\documentclass[12pt]{iopart}
\usepackage{psfig}
\usepackage{iopams}
\newcommand{\bey}[1]{\begin{eqnarray} \label{#1}}
\newcommand{\eey}{\end{eqnarray}}
\newcommand{\beq}[1]{\begin{equation} \label{#1}}
\newcommand{\eeq}{\end{equation}}

\begin{document}

\jl{1}

\title{``Superfast'' Reaction in Turbulent Flow with Potential Disorder}

\author{
Nga le Tran,
Jeong-Man Park\footnote{Permanent address:
Department of Physics, The Catholic University, Seoul, Korea}, and
Michael W. Deem
}

\address{
Chemical Engineering Department, University of
California, Los Angeles, CA  90095-1592 USA}

\begin{abstract}
We explore the regime of ``superfast'' reactivity that has been
predicted to occur in turbulent flow
in the presence of potential disorder.  Computer simulation studies
confirm qualitative features of the previous renormalization group 
predictions, which were 
based on a static model of turbulence.  New
renormalization group calculations
for a more realistic, dynamic model of turbulence  show that the
superfast regime persists.  This regime,
with concentration decay exponents greater than
that for a well-mixed reaction, appears to be a general result of the
interplay among non-linear reaction kinetics, turbulent transport,
and local trapping by potential disorder.

\end{abstract}

\submitted
\maketitle


\section{Introduction}
Reactive turbulent flow is important in a variety of natural
processes, ranging from the production of smog in the
atmosphere\cite{Bilger} to the feeding habits of certain oceanic 
creatures\cite{Kiorboe}.  The mixing effects of turbulence are put to
use in a variety of engineering applications, including combustion
reactors, fluid catalytic cracking units, and polymerization
reactors.  The behavior of reactive turbulent flow is usually
analyzed, numerically or analytically, with the continuum
reaction/transport equations \cite{Bilger,Hill}.  These mean-field
equations fail, however, in two dimensions at low densities of
reactants.  This is understood in a general way,
since the upper critical dimension for bimolecular reactions,
even in the absence of turbulence, is two \cite{Peliti,Lee2,Deem1,Deem2}.

A renormalization group treatment of a  model of
reactive turbulent flow has recently predicted a regime of
``superfast'' reactivity for the $A + A \to \emptyset$ reaction in
two dimensions \cite{Deem4}.  The superfast reactivity occurs when a
certain degree of quenched potential disorder is added to the
turbulent system.  In the superfast regime, the reactant concentration
is predicted to decay more quickly than in a well-mixed reaction:
\emph{i.e.}\ 
$c(t) \sim a/t^{1+x}$ at long times, with $x>0$.  This rapid decay occurs
due to a subtle combination of the effects of turbulent transport,
trapping by disorder, and reaction.  It shows up in the
analytical calculations as a renormalization of the effective
reaction rate.    It was anticipated that this result may have
technological implications for certain thin-film reactors or fluidized
MEMS devices.

The renormalization group calculations were carried out with a
statistical model of turbulence, in the same spirit as in treatments
of the turbulent transport of passive scalars
\cite{Fisher,Kravtsov1,Majda2}.  A review of this approach to the
dynamics of a passive scalar can be found in \cite{Bouchaud3}.  The
statistics of the turbulent velocity field in the model
were chosen to reproduce
the correct transport properties at  long times.  There was freedom
to choose either a time-dependent or time-independent velocity field,
and the time-independent choice was made.  This choice corresponds to
G.\ I.\ Taylor's hypothesis of frozen turbulence.  

In this paper, we perform further studies to confirm the 
existence of the superfast reaction regime.
  In Section 2, we describe a
method for simulating reactive turbulent flow with random,
time-independent velocity fields.  In Section 3, we present and
discuss our computational  results.  The existence of the superfast
regime is verified,  and additional features arising from
higher-loop
contributions in the renormalization group are found.  In Section 4,
we present a renormalization group treatment of reactive
turbulent flow with a  more realistic,
time-dependent model of the
velocity field.  When the appropriate type of
potential disorder is included, the superfast regime is found to
persist.  We conclude in Section 5.

\section{Simulation Methodology}
We consider the reaction
\beq{1}
A+A
~{\mathrel{\mathop{\to}\limits^{\lambda}}}~ \emptyset
\eeq
occurring in a model of a two-dimensional, turbulent fluid.  In the
absence of reaction, the $A$ particles are advected by the fluid
streamlines.  This chaotic motion is superimposed upon the natural
Brownian motion of the particles.  In addition, the particles
experience a force due to the spatially-varying, random potential. In
the presence of reaction, two $A$ particles react with the conventional
reaction rate constant $\lambda$ when in contact.

We consider this reaction to proceed
on a square $N \times N$ lattice, both for
computational convenience and to allow a direct comparison with the
field-theoretic results.  How the configuration of reactants on the
lattice changes with time is specified by a master equation.  The
master equation defines transition rates for all of the possible
changes to a given configuration of reactants on the lattice:
\bey{2}
\fl
\frac{\partial P(\{n_i\},t)}{\partial t} = \sum_{ij}
[\tau_{ji}^{-1} (n_j +1)P(...,n_i -1,n_j +1,...,t)
- \tau_{ij}^{-1} n_i P]
\nonumber \\
+ \frac{\lambda}{2 h^2} \sum_i \left[(n_i+2)(n_i +1)
 P (...,n_i +2,...,t)
-n_i (n_i -1)P \right] \ ,
\eey
where $n_i$ is the number of $A$ particles on lattice site $i$,
and $h$ is the lattice spacing.  The rate of
hopping from lattice site $i$ to lattice site $j$ is given by
$\tau_{ij}^{-1}$, which will be defined below.  The summation over
$i$ is over all sites on the lattice, and the summation over $j$ is
over all nearest neighbors of site $i$.  The particles are initially
placed at random on the lattice, with average density $n_0$.  The
initial concentration at any given site is, therefore, a Poisson random
number:
\bey{3}
P(n_i) =
\frac{[n_0 h^2]^{n_i}}{n_i !}
{\rm e}^{-n_0 h^2}
\nonumber \\
\langle n_i /h^2 \rangle = n_0 \ .
\eey

The turbulence and potential disorder enter this master equation
through the hopping rates  $\tau_{ij}^{-1}$.  These rates are chosen
so that the master equation reduces to the conventional transport
equation when there is no reaction:
\beq{4}
\frac{\partial c}{\partial t} = D \nabla^2 c + \beta D \nabla
\cdot \left[ c \nabla u - c \nabla \times \phi\right] \ ,
\eeq
where $c(\bi{r}, t)$ is the concentration of the $A$ particles, $D$ is the
diffusion constant, $\beta = 1/(k_\mathrm{B} T)$ is the inverse
temperature, $u(\bi{r})$ 
is the random potential, and $\phi (\bi{r})$ is proportional to
the random stream function.  A particularly simple form for the
transfer rate from position $i=\bi{ r}$ to the position $j=\bi{ r} +
\Delta \bi{ r}$ is 
\beq{5}
\tau_{ij}^{-1} = \frac{D}{h^2} [ 1 - \beta \Delta \bi{ r} \cdot (\nabla u
- \nabla \times \phi)/2] \ .
\eeq
When equation (\ref{5}) is used in equation (\ref{2}), one finds via
a Taylor series expansion in $h$ that the conventional transport equation
(\ref{4}) is reproduced.  This form of the transfer rates (\ref{5}) is the
one conventionally used
in theoretical calculations \cite{Deem1}.
In our simulations we choose, instead, to use the form
\beq{6}
\tau_{ij}^{-1} = \frac{D}{h^2} \exp\{\beta(u_i - u_j)/2 \} 
\exp \{\beta (\phi_{j'} - \phi_i)/2 \} \ .
\eeq
Here $j$ and $j'$ are nearest neighbors of site $i$.  The identity of site
$j'$ is derived from site $j$ by a counter-clockwise rotation of
$\pi/2$ about site $i$.  This form leads to a transition rate in the
simulation that is
bounded and non-negative, in contrast to the simpler form of equation (\ref{5}).
A Taylor series expansion of equation (\ref{6}) leads directly to the 
simpler equation (\ref{5}).  Since both of these forms are equivalent in the
limit of a small lattice spacing, they are expected to lead
to identical scaling in the  long-time regime.  

Following the usual statistical approach to turbulence,
we take the stream function and potential to be Gaussian random variables
with correlation functions given by
\bey{7}
\hat \chi_{\phi \phi} ( \bi{ k}) = \frac{\sigma}{k^{2+y}}
\nonumber \\
\hat \chi_{u u} ( \bi{ k}) = \frac{\gamma}{k^{2+y}} \ ,
\eey
where the Fourier transform of a function in two dimensions
is given by 
$\hat f(\bi{ k}) = \int \rmd^2 \bi{ r} f(\bi{ r}) \exp( \rmi 
\bi{ k} \cdot \bi{ r})$.
The random stream function and potential are generated  first in
Fourier space, where the values at different wave vectors are
independent Gaussian random variables, and then converted to real
space with a fast Fourier transform \cite{Deem5}.  Periodic boundary
conditions are used on the lattice.  The quality of the random number
generator is of importance when simulating systems with long-ranged
correlations \cite{Deem5}.  We use a  sum of three linear
congruential generators method \cite{Byte}.

Isotropic, fully-developed turbulence is modeled by $y = 8/3$ and
$\gamma = 0$.  Developing turbulence is modeled by smaller, positive
values of $y$.  Potential disorder of the form we consider here could
arise in a reaction between ionic species confined to a thin film of
fluid between spatially-addressable electrodes or media with
quenched, charged disorder.  The ionic species would move in the
random potential generated by the electrodes or quenched disorder. 
The electrodes or quenched disorder could be devised so as to
reproduce the correlation function $ \chi_{u u}(r)$.

The master equation (\ref{2}) can be exactly solved by a Markov
Poisson process \cite{Doering}.    In such a process,
we consider the motion of discrete
$A$ particles on the lattice.  For any given configuration
of reactants at time $t$,
there are $4 \sum_i n_i$ possible hopping events.  Each occurs with a
rate given by equation (\ref{6}).  In addition, there are $\sum_i n_i
(n_i-1)/2$ possible reaction events on the lattice.  Each of these
occurs with the rate $\tau_\mathrm{rxn}^{-1} = \lambda / h^2$.

The Markov process is started by initially placing the reactants on
the lattice at random with average density $n_0$.  Each step of the
Markov process consists of randomly picking one of the possible
reaction or diffusion events and incrementing time appropriately.
The probability of event $\alpha$ occurring, out of all the possible
diffusion and reaction moves, is
\beq{8}
P({\rm event~ }\alpha) = \frac{\tau_\alpha ^{-1}}{\sum_\gamma \tau_\gamma
^{-1}} \ .
\eeq
After performing the chosen event, time is incremented by
\beq{9}
\Delta t = \frac{-\log \zeta}{\sum_\gamma \tau_\gamma^{-1}} \ ,
\eeq
where $\zeta$ is a random number uniformly distributed between 0 and 1.
The Markov chain is continued until zero or one reactants remain on
the lattice, at which point no more reaction events will occur.

This Markov process, when averaged over initial conditions and 
trials, exactly reproduces the predictions of the master equation.  
Averages over the statistics of the turbulence and potential disorder
are taken by directly averaging over sufficiently many
instances of the random fields.
Since the
asymptotic scaling regime is reached only for long times, we
make two approximations to facilitate the computations.    First, we
set the reaction rate to infinity.  That is, when a particle moves to
a site with another particle already present, the reaction occurs
immediately.  Physically, we do not expect this to modify the
long-time decay law, since at long times the reaction will be in
a transport-limited regime.  Indeed, the renormalization group prediction
for the decay exponent is independent of the reaction rate
\cite{Deem4}.  The predicted prefactor also has a well-defined value
in the infinite reaction rate limit.  Second, we assume that each
particle on the lattice is equally likely to undergo a hopping event:
\beq{10}
P({\rm moving~particle} ~ \alpha) = \frac{1}{n} \ ,
\eeq
where $n=\sum_i n_i$ is the total number of particles on the lattice.
After the particle to move is chosen, the probabilities of each of the
four possible hopping events are derived from
equation (\ref{6}).  In particular, another uniform random
number is generated  for comparison against the four different hop
probabilities:
\beq{11}
P({\rm hop} ~ i) = \frac{\tau_i^{-1}}{\sum_j \tau_j^{-1}} \ .
\eeq
After the chosen hop is performed, time is incremented by
$\Delta t = N^2/(n \sum_{ij} \tau_{ij}^{-1})$.  The uniform
choice of particles to move and the approximate time incrementation
are not expected to influence the long-time exponents.  In
particular, these
approximations are exact in the long-time limit if the hopping
rates of equation (\ref{5}) are used.

\section{Simulation Results and Discussion}
\begin{figure}[p]
\section*{Figure Captions}
\centering
\leavevmode
\psfig{file=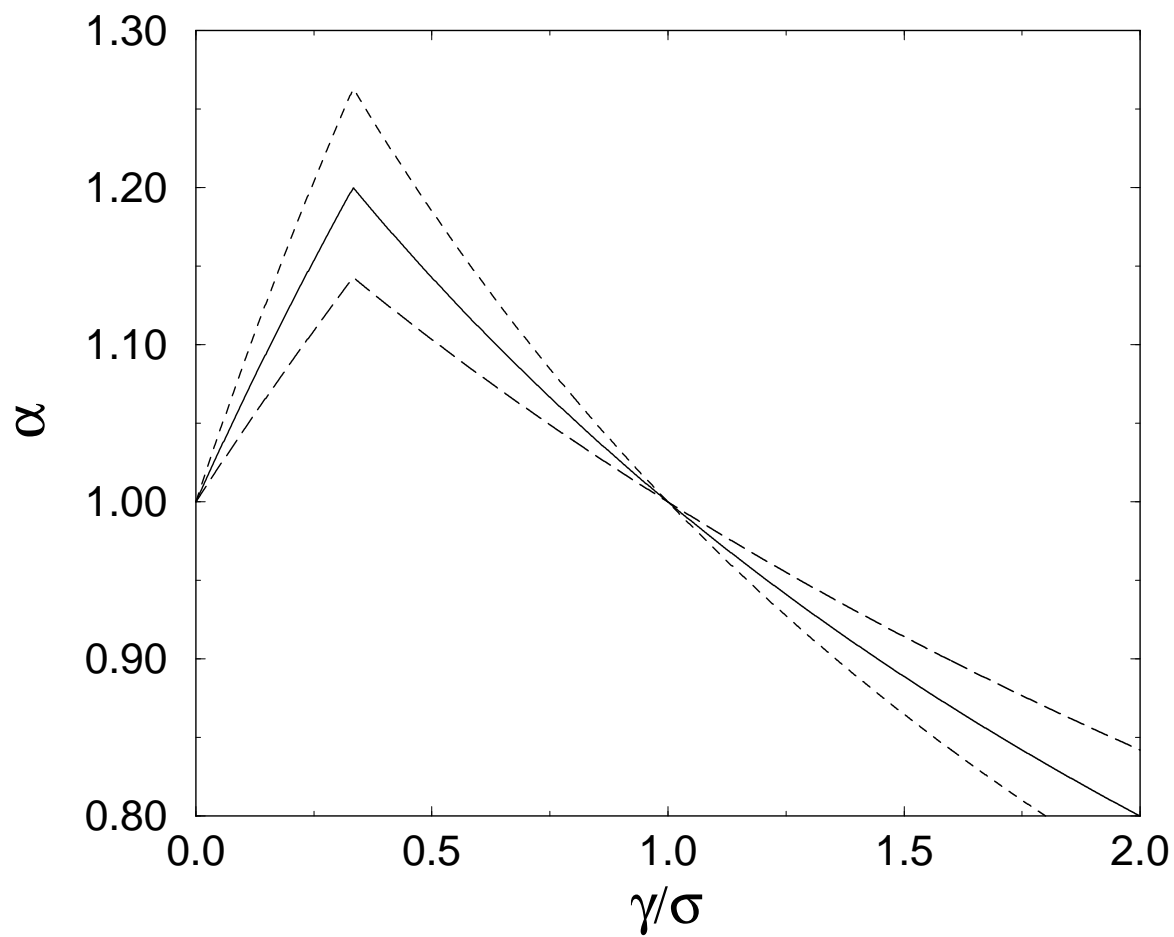,height=5in,angle=-90}
\caption[]
{\label{fig1}
The decay exponent for the
$A+A \to \emptyset$ reaction: $c(t) \sim ({\rm const}) t^{-\alpha}$.
The correspond to $y = 1$ (solid), $y= 5/4$ (short-dashed), and
$y=3/4$ (long-dashed).
}
\end{figure}

The renormalization group calculations make a prediction for the
concentration decay exponent, the $\alpha$ in $c(t) \sim a
t^{-\alpha}$ \cite{Deem4}.  The predictions depend on the properties
of the turbulence and potential disorder through the
parameters $y$, $\sigma$, and
$\gamma$.  The predictions for several values of
$y$ are shown in figure \ref{fig1}. 
The decay exponent initially rises with increasing potential disorder
and eventually decreases.  The maximum reaction rate of $\alpha = 1 +
y/(6-y)$ occurs for $\sigma = 3 \gamma$.  The decay rate depends on
$\sigma$ and $\gamma$ only through the combination $\gamma/\sigma$. 
These predictions come from a one-loop renormalization group
calculation, and they are strictly valid only for small $y$.  In
addition, the predictions
 should be more accurate for small $\gamma$, because it
is known that higher-loop corrections lead to a modification of the
flow diagram for large $\gamma$.

We find that our Poisson process efficiently solves the master
equation for values of $y$ near unity.  For smaller values of $y$,
the renormalization of the effective reaction rate occurs 
slowly.  This means that the predicted asymptotic scaling occurs
only at long times, times longer than we can reach in our
simulation.  For larger values of $y$, significant lattice effects
occur, due to our choice of transition rates, equation  (\ref{6}), and
correlation functions, equation  (\ref{7}).

\begin{figure}[p]
\centering
\leavevmode
\psfig{file=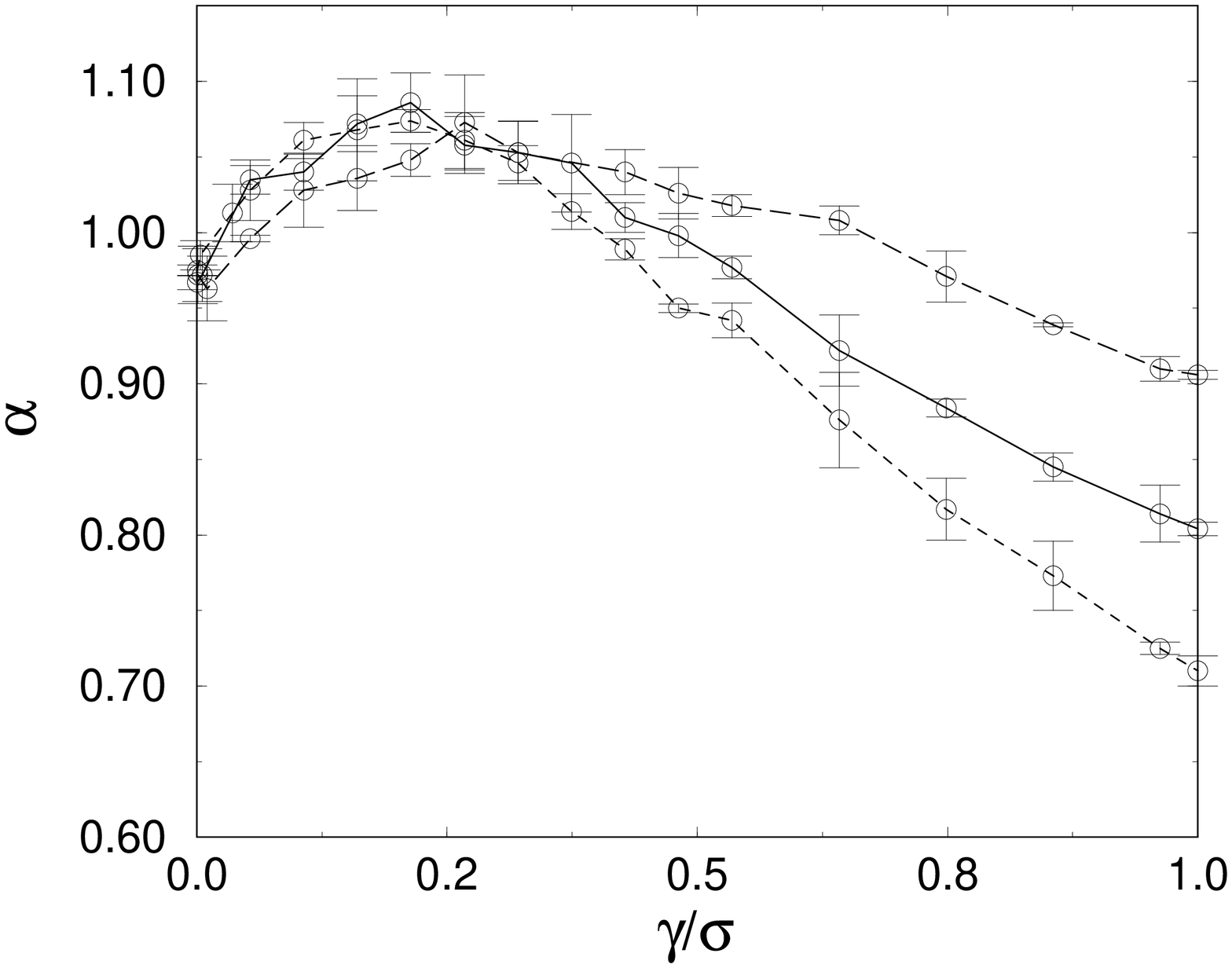,height=5in,angle=-90}
\caption[]
{\label{fig2}
The decay exponent observed in the simulation for
$y=3/4$.
The three curves correspond to $\sigma = \sigma^*$ (solid),
$\sigma = 3 \sigma^*/2$ (short-dashed), and $\sigma = \sigma^*/2$
(long-dashed).
}
\end{figure}

\begin{figure}[p]
\centering
\leavevmode
\psfig{file=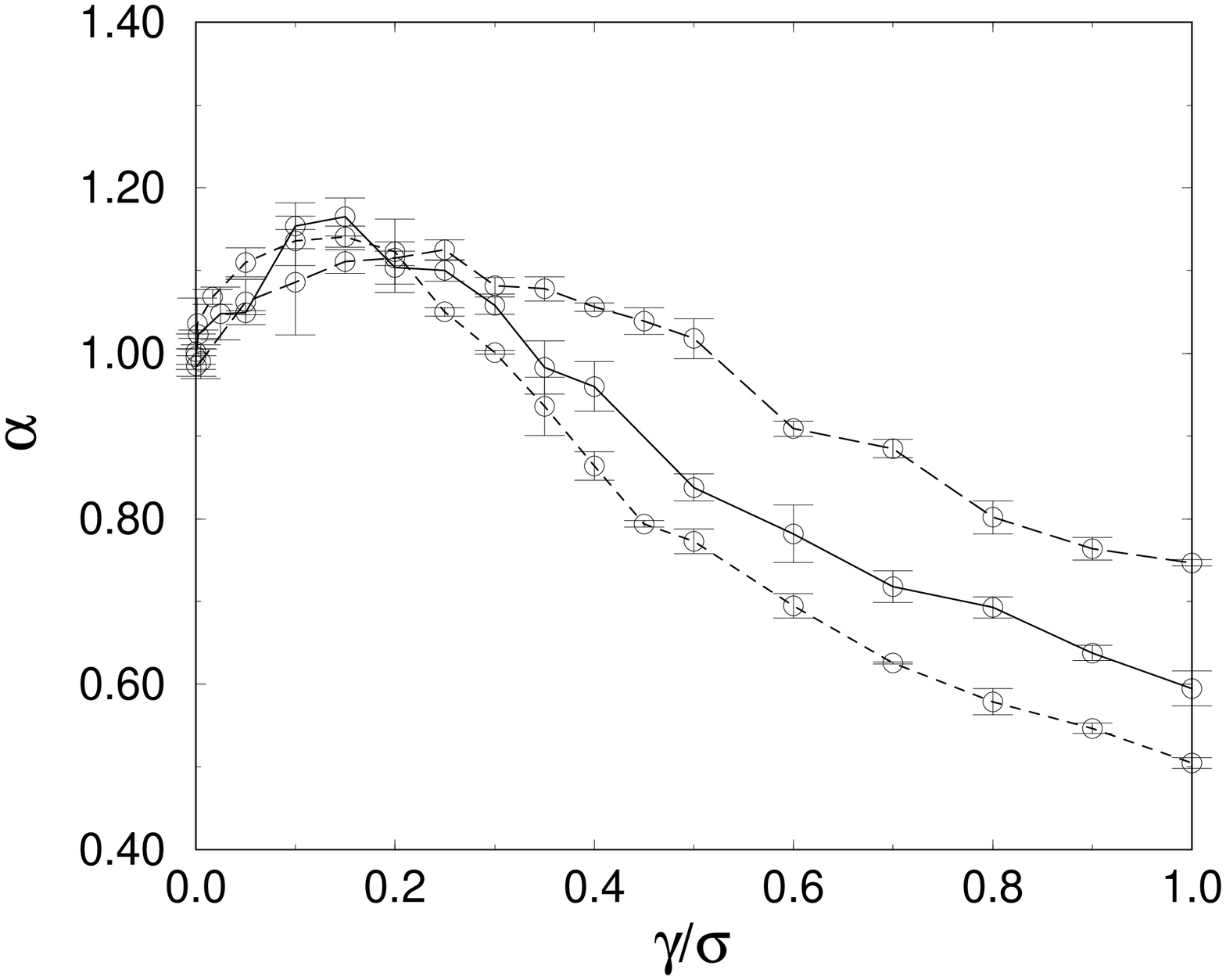,height=5in,angle=-90}
\caption[]
{\label{fig3}
The decay exponent observed in the simulation for
$y=1$.
The three curves correspond to $\sigma = \sigma^*$ (solid),
$\sigma = 3 \sigma^*/2$ (short-dashed), and $\sigma = \sigma^*/2$
(long-dashed).
}
\end{figure}

\begin{figure}[p]
\centering
\leavevmode
\psfig{file=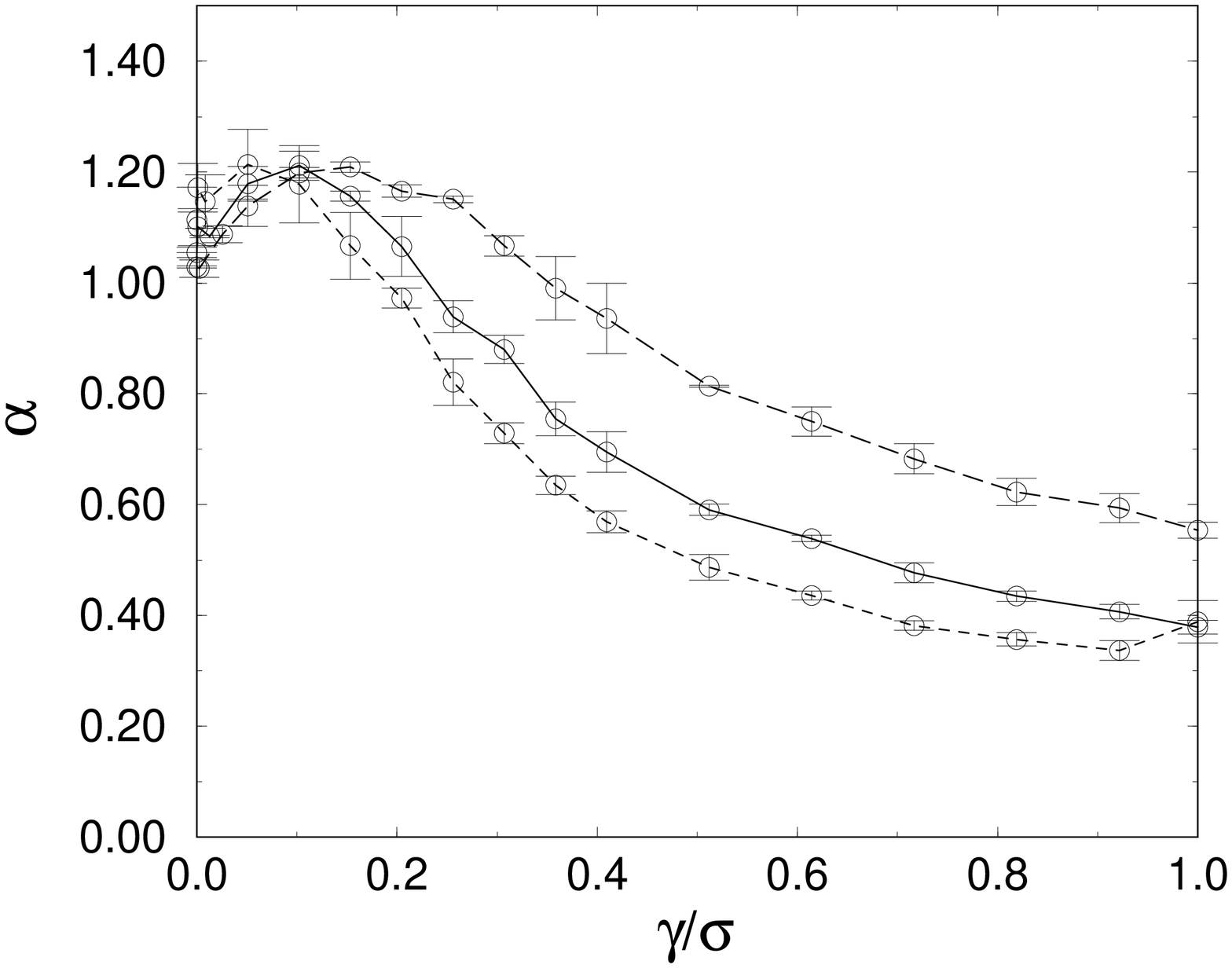,height=5in,angle=-90}
\caption[]
{\label{fig4}
The decay exponent observed in the simulation for
$y=5/4$.
The three curves correspond to $\sigma = \sigma^*$ (solid),
$\sigma = 3 \sigma^*/2$ (short-dashed), and $\sigma = \sigma^*/2$
(long-dashed).
}
\end{figure}

Shown in figure \ref{fig2} are the decay exponents observed in our
simulation for $y=3/4$.  The decay exponents observed for $y=1$ and
$y=5/4$ are shown in figures \ref{fig3} and \ref{fig4},
respectively.  Each of the data points in the figure is an average
over three different runs on three different instances of the
turbulence and potential disorder.  The standard deviations estimated
from these three runs are shown as the error bars.  

\begin{figure}[p]
\centering
\leavevmode
\psfig{file=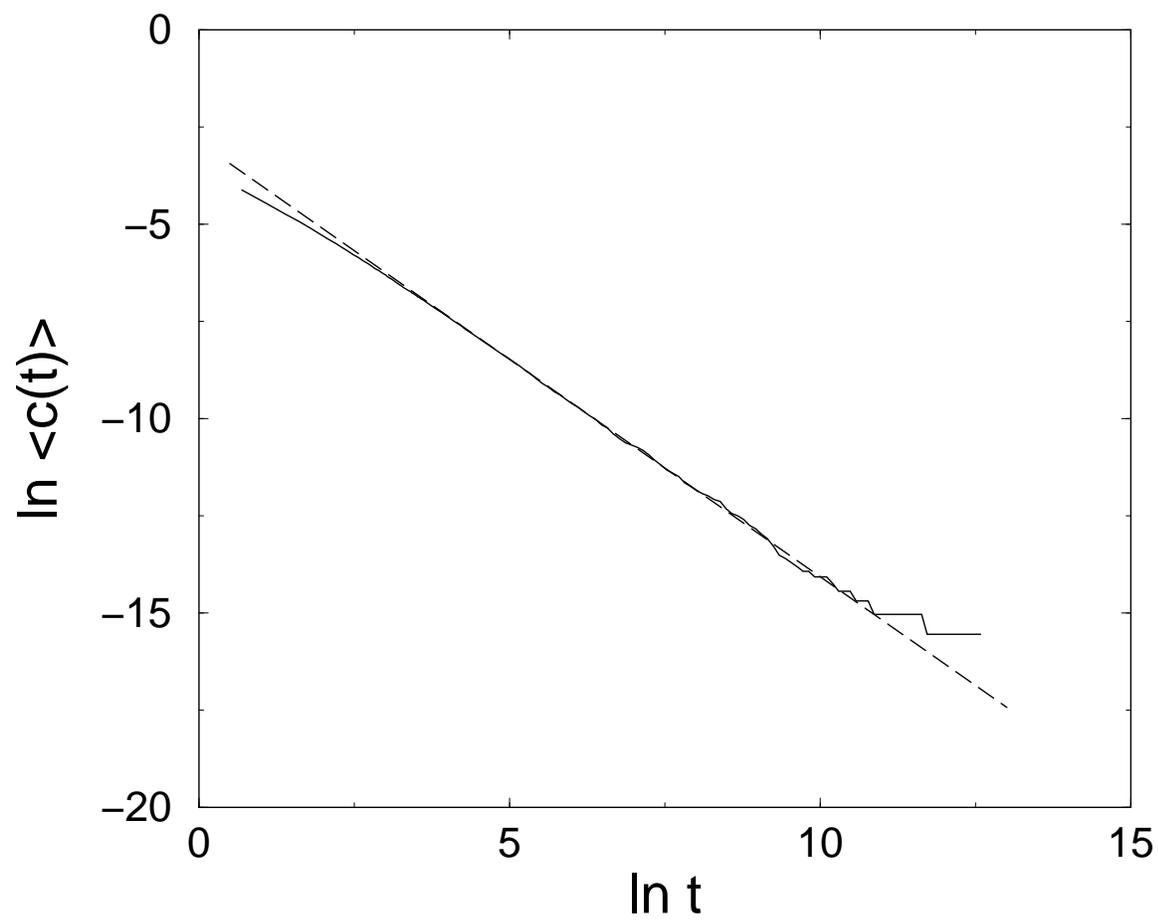,height=5in,angle=-90}
\caption[]
{\label{fig6}
The concentration as a function of time (solid line) for the case $y=1$, 
$\sigma =  \sigma^*$, and $\gamma/\sigma = 0.2$.
Also shown is the fit to the data in the scaling regime (dashed line).
}
\end{figure}

We made a few canonical choices for constants in our simulation.  So
as to reach the asymptotic scaling regime, we used $4096 \times 4096$
square lattices.  Simulations on $2048 \times 2048$ lattices gave
similar results.  
Finite size effects should be most noticeable for small values of
 $y$, $\sigma$, or $\gamma$, since in that regime the renormalization of
the effective reaction rate is most slow.  Indeed,
finite size effects only appear to be present
in figure \ref{fig2} for small $\gamma/\sigma$, where
the slow renormalization of $\lambda(l)$ leads to an observed
decay exponent that is below the correct asymptotic value of unity.
While additional finite size effects could be present, the agreement
at small $\gamma/\sigma$ among
 curves for different $\sigma$ but same $\gamma/\sigma$ and $y$, and the
agreement between the results for $2048 \times 2048$ and
$4096 \times 4096$ lattices, argues against this.
We used a lattice spacing of $h=1$, which we can
enforce by a spatial rescaling, and a diffusion constant of $D=1$,
which we can enforce by a temporal rescaling.  The long-time decay
exponent is not affected by these rescalings.  We collected the
time-dependent concentration data on the lattice by binning the
results into a
histogram with a temporal bin width of $\delta t = 1$.  We continued each
simulation until the concentration was so low, $c \approx
(\mathrm{const}) /N^2$, that lattice effects were obvious.
We show in figure \ref{fig6} the simulation results for a typical run.
The average of the slopes determined in three such runs
gives one of the data points in figures \ref{2}-\ref{4}.

Since the renormalization group predictions depend only on
$\gamma/\sigma$, we fixed $\sigma$ and varied $\gamma$ for each
simulation at a fixed value of $y$.  Under these conditions, the
effective strength of turbulence is expected to flow to its fixed
point value, $\sigma \to \sigma^* \approx 2 \pi y (2 \pi / h)^y$ \cite{Deem4}.
  In each
of the figures \ref{fig2}-\ref{fig4} we show results for three different
``bare''
values of $\sigma$: $\sigma=\sigma^*$, $\sigma=\sigma^*/2$, 
and $\sigma=3 \sigma^*/2$.

We use a $\ln c(t)$ \emph{versus} $\ln t$ plot to obtain the
decay exponent.  This plot has many more data at large $\ln t$,
 which leads to an over-weighting of the long-time
regime.  To counteract this effect, we use data exponentially spaced
in time.  We exclude both short-time data, which are not in the
asymptotic regime, and long-time data, which show finite-size
effects.  Use of exponentially spaced data allows one to obtain a
reliable estimate of the error in the exponent determined for a given
realization of the disorder.  We find that this statistical error is
smaller than the systematic deviation that occurs for
each different realization of the disordered streamlines
and potential.  The error bars shown in the
figures encompass both the statistical and the systematic errors.

We see that there is a general agreement between the simulation
results and the renormalization group predictions.  In particular, 
the superfast reaction regime, $\alpha > 1$, is observed.  This was a
dramatic analytical prediction, and confirmation of this regime is
significant.  The exponent is observed to
reach roughly the maximum value
of $1 + y/(6-y)$ predicted by theory.

The agreement between the simulation results and the one-loop
renormalization group predictions is better for smaller $y$.
 For small $y$, the location of the peak in reactivity is
trending towards the predicted value of $\gamma/\sigma = 1/3$.  For
finite values of $y$, the location of the peak is shifted to the
left.  Indeed, the whole curve is shifted to the left for finite $y$.
One consequence of this is that the observed decay exponent decreases below
unity for $\gamma/\sigma$ close to, but less, than one, in contrast to the
one-loop predictions.

The agreement between the simulation results and the one-loop
predictions is also better for small $\gamma$.  For small
$\gamma/\sigma$, the simulation results for different values of
$\sigma$ fall on the same curve.  This is the universal curve
predicted at one-loop, albeit compressed to the left.
  For large $\gamma$, the simulation results
depend on both $\gamma$ and $\sigma$, not simply on the ratio
$\gamma/\sigma$.  This dependence is presumably due to corrections that 
would enter in a higher-loop calculation.

\section{Renormalization Group Treatment of Dynamic Turbulence}

The simulations, and the previous renormalization group calculations,
were performed for a model of turbulence with streamlines random in
space, but constant in time.  For this model, both the analytical and
computational studies predict a regime of superfast reaction in the
presence of potential disorder.  This regime was understood to occur
as a general consequence of the interplay among turbulent transport,
trapping by the random potential, and a non-linear dependence of the
reaction rate on local reactant concentration.  This regime was not
believed to occur, for example, due to some subtle, artificial interaction
between the quenched potential and quenched streamlines.

Even so, it would be interesting to predict the existence of the
superfast reaction regime for a dynamic model of turbulence.  As we
have seen, reliable simulations require large lattices and long
times, even for quenched turbulence.  Moreover, simulations were
possible only for an intermediate range of $y$, due to the competing
considerations of minimizing lattice effects and  accessing
the asymptotic regime.  Renormalization group theory is,
however, a viable approach to studying
reactive turbulent flow in a dynamic model of turbulence.

We, again, use a statistical model of turbulence \cite{Majda2}.
We now assume that the stream function is random in both space
and time:
\beq{12}
\hat \chi_{\phi \phi} ( \bi{ k}, t_1-t_2) = \frac{\sigma}{k^{2+y_\sigma}}
\vert t_1 - t_2 \vert ^{-\rho} \Theta(t_1-t_2) \ ,
\eeq
where $\Theta(t)$ is the Heavyside step function.  A family of models
for turbulence is generated by varying $\rho$  and $y_\sigma$. 
Isotropic turbulence is modeled when $3 y_\sigma - 2 \rho = 8$, at
least for small $\rho$.  We include a random potential, as before:
\beq{12a}
\hat \chi_{u u} ( \bi{ k}) = \frac{\gamma}{k^{2+y_\gamma}} \ .
\eeq
We will find that the most interesting regime occurs for $y_\gamma
< y_\sigma$, due to the weakening of the random stream function
through decorrelation over time.

As in previous studies \cite{Deem4}, we map the master equation
(\ref{2}) onto a field theory using the coherent state representation
\cite{Peliti,Lee1}.  For this operation, we use the transition rates
given by equation (\ref{5}).  The random and stream function are
incorporated with the replica trick \cite{Kravtsov1}, using $N$
replicas of the original problem.  The concentration of reactants at
time $t$, averaged over the initial conditions, $c(\bi{ r},t)$, is
given by
\beq{13}
c(\bi{ r},t) = \lim_{N \to 0} \langle a(\bi{ r},t) \rangle \ ,
\eeq
where the average is taken with respect to  $\exp(-S)$, with
\bey{14}
\lo{S =} \sum_{\alpha=1}^N
\int \rmd^2 \bi{ r} \int_0^{t_\mathrm{f}} \rmd t
 \bar a_\alpha(\bi{ r},t) \left[
\partial_t - D \nabla^2 + \delta(t)
 \right]
 a_\alpha(\bi{ r},t)
 \nonumber \\
+\frac{\lambda}{2}  \sum_{\alpha=1}^N
\int \rmd^2 \bi{ r} \int_0^{t_\mathrm{f}} \rmd t \bigg[
2 \bar a_\alpha(\bi{ r},t)
 a_\alpha^2(\bi{ r},t)
+ \bar a_\alpha^2(\bi{ r},t)
 a_\alpha^2(\bi{ r},t)
\bigg]
\nonumber \\ 
  -n_0  \sum_{\alpha=1}^N \int \rmd^2 \bi{ r} \bar a_\alpha(\bi{ r},0)
\nonumber \\
 -\frac{\beta^2 D^2}{2}
 \sum_{\alpha_1,\alpha_2=1}^N
\int \rmd t_1 d t_2 \int_{\bi{ k}_1 \bi{ k}_2 \bi{ k}_3 \bi{ k}_4}
 (2 \pi)^2 \delta(\bi{ k}_1+\bi{ k}_2+\bi{ k}_3+\bi{ k}_4)
\nonumber \\ 
\times
\hat{\bar a}_{\alpha_1}(\bi{ k}_1, t_1)
\hat{     a}_{\alpha_1}(\bi{ k}_2, t_1)
\hat{\bar a}_{\alpha_2}(\bi{ k}_3, t_2)
\hat{     a}_{\alpha_2}(\bi{ k}_4, t_2)
\nonumber \\ 
\times \bigg[
\bi{ k}_1 \cdot (\bi{ k}_1+\bi{ k}_2)
\bi{ k}_3 \cdot (\bi{ k}_3+\bi{ k}_4)
\hat\chi_{uu}(\vert \bi{ k}_1+\bi{ k}_2\vert)
\nonumber \\
+
\bi{ k}_1 \times \bi{ k}_2~
\bi{ k}_3 \times \bi{ k}_4
\hat\chi_{\phi \phi}(\vert \bi{ k}_1+\bi{ k}_2\vert, t_2-t_1) \bigg]  \ .
\eey
The notation
$\int_\bi{ k}$ stands for $\int \rmd^2 \bi{ k} / (2 \pi)^2$.
The upper time limit in the action is required only to satisfy
$t_\mathrm{f} \ge t$.

Using renormalization group theory, we can derive the flow equations
for this model.  We find that the turbulence contributes directly to
the dynamical exponent, and therefore to the transport properties,
 but not directly
to the effective reaction rate.  The potential disorder contributes
directly to both the dynamical exponent and the effective reaction
rate.  Finally, the dynamical exponent contributes indirectly to the effective
reaction rate as well.
We define $g_\sigma =
\beta^2 \sigma \Lambda ^{2\rho  - y_\sigma}D^\rho \Gamma(1-\rho)/ (4 \pi)$
and
$g_\gamma = \beta^2 \gamma \Lambda^{- y_\gamma} / (4 \pi)$,
where $\Lambda = 2 \pi /h$ is the cutoff in Fourier space, and
$\Gamma(x)$ is the standard Gamma function.  The precise definition
of $g_\sigma$ depends on the regularization of $\chi_{\phi \phi}$
at $t=0$.
At one loop, we find the flow equations to be
\bey{15}
\frac{d \ln n_0}{d l} = 2
\nonumber \\
\frac{d \ln \lambda}{d l} = - 
   \frac{\lambda}{4 \pi D} +3 g_\gamma - g_\sigma 
\nonumber \\
\frac{d \ln g_\gamma}{d l} = y_\gamma  - 2 g_\sigma
\nonumber \\
\frac{d \ln g_\sigma}{d l} = y_\sigma - \rho z  - 2 g_\sigma \ .
\eey
The dynamical exponent is given by
\beq{16}
z = 2  +  g_\gamma- g_\sigma  \ .
\eeq

To achieve the superfast regime, we must set 
$y_\gamma = y_\sigma - \rho z$.
Without this choice, the potential disorder will either dominate or be
irrelevant.  It is natural that the appropriate
$y_\gamma$ should be weaker than
$y_\sigma$, since the effective turbulence strength is weakened by
decorrelation effects over time.  With this choice, we find that the
ratio $g_\gamma(l)/g_\sigma(l)$ remains constant under the
renormalization group flows.  The parameter $g_\sigma(l)$, however,
flows to the fixed point $g_\sigma^* = y_\gamma/2$.  Moreover, we
find that when $g_\gamma/g_\sigma > 1/3$, the effective reaction rate
flows to a finite fixed point $\lambda^* = 4 \pi D(3 g_\gamma^* -
g_\sigma^*)$.  When $g_\gamma/g_\sigma < 1/3$, the effective reaction rate
renormalizes to zero.

With the choice of
$y_\gamma = y_\sigma - \rho z$,
 the flow equations are
the same as for the static model of turbulence \cite{Deem4}.  The
matching is the same, as well.  In terms of these new variables,
therefore, the predicted concentration dependence is the same. 
\emph{In particular, the superfast reaction regime is predicted to
exist for this dynamic model of turbulence.} The explicit, long-time
concentration dependence  for weak disorder is
\bey{18}
c(t) \sim \left[ \frac{1}{4 \pi D (g_\sigma^* - 3 g_\gamma^*)}
 + \frac{1}{\lambda_0} \right] \frac{1}{t}
\left( \frac{t}{t_0} \right)^{-
 2 g_\gamma^*/(2 + g_\gamma^* - g_\sigma^*)} , ~~~(3 g_\gamma < g_\sigma) 
\ ,
\eey
where $\lambda_0$ is the bare value of the reaction
rate.  The matching time$, t_0$, is given roughly by
$t_0 \approx h^2/(2 D)$.  For strong disorder, the concentration decays as
\bey{19}
c(t) \sim  \frac{1}{4 \pi D(3 g_\gamma^* - g_\sigma^*) t}
\left(\frac{t}{t_0}\right)^{
(g_\gamma^* - g_\sigma^*)/(2 + g_\gamma^* - g_\sigma^*)} , 
~~~(3 g_\gamma > g_\sigma)  \ .
\eey
At the location of the maximum reaction rate, there is a
logarithmic correction to power-law decay:
\beq{20}
c(t) \sim  \frac{\ln (t/t_0)}{8 \pi (1 -y_\gamma/6) D t}
t^{-y_\gamma/(6-y_\gamma)} 
, ~~~(3 g_\gamma = g_\sigma)  \ .
\eeq

\section{Conclusion}
A regime of reaction rates faster than that for a well-mixed reaction
was observed in simulations of a static model of turbulence plus
potential disorder.  Qualitative predictions of previous
renormalization group studies of this model were reproduced. 
Interesting departures from the one-loop predictions, such as a
compression of the decay exponent curve to the left, were observed. 
These features are expected to be reproduced in higher-loop
analytical calculations.  

Renormalization group calculations for a more
realistic, dynamic model of turbulence show that the superfast regime
persists.    Indeed, these calculations suggest that our general
understanding of the phenomenon of superfast reactivity is correct.  The
random potential tends to attract reactants to localized regions in
space, and by doing so increases the reaction rate in a non-linear
manner in these regions.  The reactants, therefore, annihilate at a
faster than average rate in these regions.  The turbulent velocity
fields transport reactants to these regions rapidly, although the
transport is slowed by the ruggedness of the random potential
landscape.  The net effect of the trapping and the turbulent transport is
superfast reactivity whenever the mean-square displacement is
super-linear.

The existence of a superfast  reaction regime is of great interest
from a practical point of view.  Often, it is assumed in reactor design
that the highest reaction rate is obtained for a well-mixed system
\cite{Jensen}.  Our results show that under certain conditions,
the highest reaction rate is achieved, instead, for an inhomogeneous
system.  This behavior has been observed experimentally.
Indeed, 
an enhancement due to inhomogeneous reactant concentrations
 of the effective reaction rate between ions
in a chaotically-mixed,
two-dimensional,
fluid with attractors has recently been observed
\cite{Tabeling}.
The superfast regime may be relevant to a variety of reactors,
including thin-film reactors, certain combustion reactors, and
certain types of micro-electro-mechanical (MEMS) devices.

\ack
This research was supported by the National Science Foundation
through grants CHE--9705165 and CTS--9702403.
\bigskip

\bibliography{nga}


\clearpage
\newpage

\end{document}